\documentclass[prc,aps,amsfonts,amsmath,superscriptaddress,floatfix,twocolumn]{revtex4}
\usepackage{graphicx,float} 
\usepackage{epsfig} 
\usepackage{rotating}
\usepackage{mwe}    
\usepackage{subfig}
\begin{document}

\title{Nucleonic localisation and alpha radioactivity}

\author{J.-P. Ebran}
\affiliation{CEA,DAM,DIF, F-91297 Arpajon, France}
\affiliation{Universit\'e Paris-Saclay, CEA, Laboratoire Mati\`ere en Conditions Extr\^emes, 91680, Bruy\`eres-le-Ch\^atel, France}
\author{E. Khan}
\affiliation{IJCLab, Universit\'e Paris-Saclay, CNRS/IN2P3, 91405 Orsay Cedex, France}
\author{ R.-D Lasseri}
\affiliation{IJCLab, Universit\'e Paris-Saclay, CNRS/IN2P3, 91405 Orsay Cedex, France}
\affiliation{ESNT, CEA, IRFU, D\'epartement de Physique Nucl\'eaire, Universit\'e Paris-Saclay, F-91191 Gif-sur-Yvette}

\begin{abstract} 
Relativistic energy density functional approaches are known to well describe nuclear states which involve alpha clusters. 
Here, alpha emitting nuclei are analysed through the behavior of the spatial localisation of nucleonic states, calculated with an axially deformed RHB approach over the nuclear chart. 
The systematic occurrence of more localised valence states, having a $n=1$ radial quantum number, allows to pinpoint nuclei in agreement with experimentally known alpha-emitters.
The cases of $^{212}$Po and $^{104}$Te are investigated, showing the concomitant contributions of the pseudospin symmetry and the presence of $n=1$ states, on the alpha preformation probability. The impact of the localisation of valence states, on alpha preformation probability, is then analysed. It allows to study shell effects on this probability, over isotopic and isotonic chains.  Finally, a phenomenological law  is also provided, relating this probability to the radial quantum number of the valence states.
\end{abstract}
 


\date{\today}

\maketitle

\section{Introduction}

The description of alpha radioactivity is a long standing problem. For about one century, a large variety of models have been devoted to this task, such as semi-classical approaches, or microscopic ones \cite{del10}. However, the identification of all nuclei, which are alpha-emitter, may not be achieved yet, as proven by the discovery of alpha-emission of $^{209}$Bi in 2003 \cite{mar03}, or the remaining question of the possible alpha radioactivity of $^{208}$Pb \cite{bee13}. One of the first successful model is the well-known Geiger-Nuttall law \cite{gei11}, and its description involving tunneling effect by Gamow \cite{gam28} and the class of WKB models  \cite{del10}. In these approaches, the alpha emission probability is decomposed into the product of an alpha preformation one with the tunneling probability through the Coulomb barrier, taking into account the frequency of impact of the alpha particle on the barrier. 

The probability of tunneling effect depends on the Q-value. The study of the alpha preformation probability itself, has received significant attention since several decades \cite{del10,lov98,del02,cho19}, although it has also often been approximated to 1. In addition, some points remain to be clarified, as first discussed by Buck and collaborators \cite{buc90}, such as the difference of behavior of the alpha emission probability for N smaller or larger than 126, respectively. This was first analysed by introducing a global quantum number in an alpha plus core description of the system, namely the Wildermuth condition \cite{buc89,buc91}. It points towards the possible role of quantum numbers in alpha radioactivity. It shall be relevant to further provide a description of this shell effect at the nucleonic scale, involving nucleon quantum numbers. Relating this effect to nuclear structure, and studying the possible impact of the alpha preformation probability, would also be of interest, as already pointed out in Refs. \cite{cho10,cho14}.

The study of the discrepancy between an accurate phenomenological law (i.e. an enhanced Geiger-Nuttall law) \cite{bro92} assuming an alpha preformation probability of 1, and the experimental alpha emission probability, also brought relevant investigations: the observed patterns, compared to experiment, are imaging the preformation probability. For instance, the N=126 shell effect has an impact on this probability  \cite{bro92}, as well as the Z=82 one, as shown more recently \cite{and13}. Considering a large variety of alpha-emitters, a variation up to a factor 30 of the preformation probability can be inferred, in the N=126 case \cite{bro92}. 

More recently, the alpha emission probability was deduced from the measurement of the alpha emission lifetime  in the $^{104}$Te nucleus \cite{aur18,xia19}, showing a larger emission probability than in $^{212}$Po, which remains to be fully explained.

Nuclear structure properties could therefore help to better understand the alpha preformation probability. Indeed, light nuclei are known to exhibit alpha cluster states. They are of course not alpha-emitters (except for $^8$Be), because of their negative Q-value. However, the occurrence of alpha cluster states, and the alpha preformation probability in heavier nuclei shall be closely related. A recent work has for example showed how to describe alpha emission in $^{104}$Te and alpha cluster states in $^{20}$Ne on the same ground, using and alpha+core approach \cite{ibr19}. $^{212}$Po alpha decay was also described with a significant alpha+core contribution \cite{varg92}.
Hence, a description of alpha emission at the fully nucleonic level could be also interesting.

In a previous work, we showed that the so-called localisation parameter, at the nucleonic scale, was driving the occurrence of cluster states \cite{ebr18}:  the radial quantum number $n$ is the key quantity impacting spatial localisation of nucleons, and hence, cluster occurrence in nuclei. Therefore, nuclei with spatially localised valence states should appear throughout the nuclear chart as favored alpha-emitters, providing a universal and simple approach: a first link with alpha radioactivity was explored in this work \cite{ebr18}. 
In the present work, we propose to further investigate links between alpha cluster formation (driven by localisation) in nuclei, and the alpha preformation probability, using the criterion of the localisation of the nucleonic wave function. The aim is twofold: i)  study if localised valence nucleonic states allow to pinpoint alpha decay emitters, and ii) analyse if the alpha preformation factor is favored by the occurrence of such states. Ultimately, a simple phenomenological law shall be provided, relating the alpha preformation factor to the radial quantum number of the valence state of the considered nucleus.

In section II, we first define a criterion for localised nucleonic states. It is then used to predict alpha-emitters over the nuclear chart, by calculating the spatial dispersion of valence states with the microscopic relativistic Hartree-Bogoliubov (RHB) approach: such a class of models has proven to be a relevant tool, to analyse cluster occurrence in nuclei on a general ground \cite{ebr12,ebr13,ebr14,ebr14a,zha15,zho16}, as well as to provide a sound comparison with measured excited spectra of cluster states, such as in Ne isotopes \cite{mar18} and for the $^{12}$C Hoyle state \cite{mar19}. It should be noted that it was also recently shown to describe alpha radioactivity of $^{104}$Te  and $^{108}$Xe \cite{mer20}. In section III, the role of localisation is analysed on the $^{104}$Te and $^{212}$Po benchmark cases, through their single-particle spectra. Section IV provides a study of the alpha preformation factor, related to nucleonic localisation. The shell effects are also analysed in this framework. A phenomenological law, relating the radial quantum number to the alpha preformation factor, is finally given.

\section{Alpha-emitters over the nuclear chart}


In order to predict the general behavior of alpha emitting nuclei over the nuclear chart, microscopic energy density functional calculations are performed. Namely, the fully self-consistent relativistic Hartree-Bogoliubov approach is used with the DD-ME2 functional for axially symmetric nuclei \cite{vre05}. Such an approach has proven to be successful to describe on the same ground a large variety of phenomena in nuclei \cite{vre05}, and the occurrence of cluster states \cite{ebr12,ebr13,ebr14,ebr14a,zha15,zho16,mar18,mar19}. The pairing interaction used in the RHB calculation is separable in momentum space, and driven by the bell-shape pairing gap in symmetric nuclear matter \cite{nik10}. 

Limitations of such an approach shall also be discussed. Regions of the nuclear chart involve nuclei with triaxiality or octupole degrees of deformation, which are not taken into account here. Moreover, drip-line nuclei may also be delicate to describe with the present approach. Finally, the Wigner term, as well as proton-neutron pairing, could have an effect, especially for N=Z nuclei, but are not considered here. It should be noted that, more generally, quartetting is also known to have an impact on clustering \cite{las18}. All these effects can be relevant, but the present approach does not aim to provide a fully detailed prediction of all the alpha-emitters over the nuclear chart. It rather aims to investigate if the spatial localisation of valence states increases alpha preformation probability.

It order to study the possible link between alpha-emitters and localised nucleonic valence states, it is first necessary to define a criterion for such a localisation. In a second step, the corresponding nuclei fulfilling this criterion (calculated with the microscopic RHB approach), shall be compared to the experimentally known alpha-emitters.

\subsection{Criterion for localised states}

A first analysis of the occurrence of cluster states can be investigated from the spatial localisation of the nucleonic degree of freedom, through the localisation parameter, defined as \cite{ebr12,ebr13,ebr18}:

\begin{equation}
\alpha_{loc}=\frac{2\Delta r}{r_0} 
\label{eq:def}
\end{equation}

where  $\Delta r = \sqrt{<r^2> - <r>^2}$ is the typical spatial dispersion of the nucleonic wave function, and ${r_0 \simeq 1.25}$ fm the typical inter-nucleon distance 
determined by nuclear saturation density ($\rho \simeq 0.16$ fm$^{-3})$. 

In order to better understand the role of the spatial dispersion, it has been shown, in the spherical harmonic oscillator approximation, that the dispersion of a given single-particle state is, to a good approximation, only driven by its radial quantum number $n$, and not the orbital one \cite{ebr18}:
\begin{equation}
\alpha_{loc}\simeq\frac{\sqrt{\hbar (2n-1)}}{(2mV_0r_0^2)^{1/4}}~A^{1/6}\;
\label{eq:res}
\end{equation}

where V$_0$ is the depth of the confining potential of the considered nucleus, composed of A nucleons of mass m. Microscopic calculations of the dispersion (using Eq. (\ref{eq:def})) show that the smallest dispersion pattern appears for single particle states with $n=1$ \cite{ebr18}, independently from the orbital quantum number. This key point opens the possibility to pinpoint nuclei having spatially localised valence nucleons (namely $n=1$ valence states), throughout the nuclear chart. Correlations between localisation and preformation of an alpha particle, in view of its emission, can be investigated.

Eqs (\ref{eq:def}) and (\ref{eq:res}) allow to determine a criterion for the spatial dispersion of valence state to be localised, in a given nucleus. Using these two equations, in the harmonic oscillator approximation, leads to:

\begin{equation}
\Delta r .  A^{-1/6}\simeq \frac{\sqrt{\hbar r_0}}{(32mV_0)^{1/4}}.\sqrt{(2n-1)}
\label{eq:condho}
\end{equation}

The prefactor value on the r.h.s of Eq. (\ref{eq:condho}) is 0.4 fm, using a typical depth of the potential V$_0$=75 MeV \cite{ebr12}. Taking into account the larger diffusivity of the nuclear potential compared the the HO one, a factor 1.2 has to be considered \cite{ebr18}, so the prefactor is about 0.5 fm. Hence, a relevant criterion over the nuclear chart to disentangle a localised $n=1$ state ($\Delta r.A^{-1/6}\sim$0.5 fm) from a more delocalised $n=2$ state ($\Delta r.A^{-1/6}\sim$0.5 fm.$\sqrt{3}\simeq$0.9 fm) is:

\begin{equation}
\langle \Delta r \rangle . A^{-1/6}<0.7 fm
\label{eq:cond}
\end{equation}

The 0.7 fm value is therefore the condition to have a small dispersion of the considered state (i.e. a $n=1$ state) in a given nucleus, taking into account the dependency on A. 
 $<\Delta r>$ is the average of the nucleonic spatial dispersion: in practice, in a microscopic calculation where pairing effects are at work, there is not a single valence state. Therefore,
 $<\Delta r>$ is calculated as the average dispersion of the valence states, weighted by their occupation probability. Since alpha preformation is under study, the number of valence states to be considered is determined until a particle number (either neutrons or protons) of 2 is reached from the sum of the considered occupation probabilities. The resulting mean dispersion allows then to pinpoint nuclei having a small spatial dispersion, by using the criterion (\ref{eq:cond}). We have checked, using the microscopic calculations, that this condition allows to pinpoint $n=1$ states over the nuclear chart.
 
A complementary way to analyse the relevance of this criterion, is to study its impact on alpha radioactivity. This shall be done in the next subsection, through systematic calculations over the nuclear chart, in order to compare with known alpha emitting nuclei. This shall also be undertaken in the last section of the present work, by the means of a phenomenological evaluation of the alpha preformation factor.

\subsection{Calculation over the nuclear chart}

\begin{figure}[tb]
\scalebox{0.35}{\includegraphics{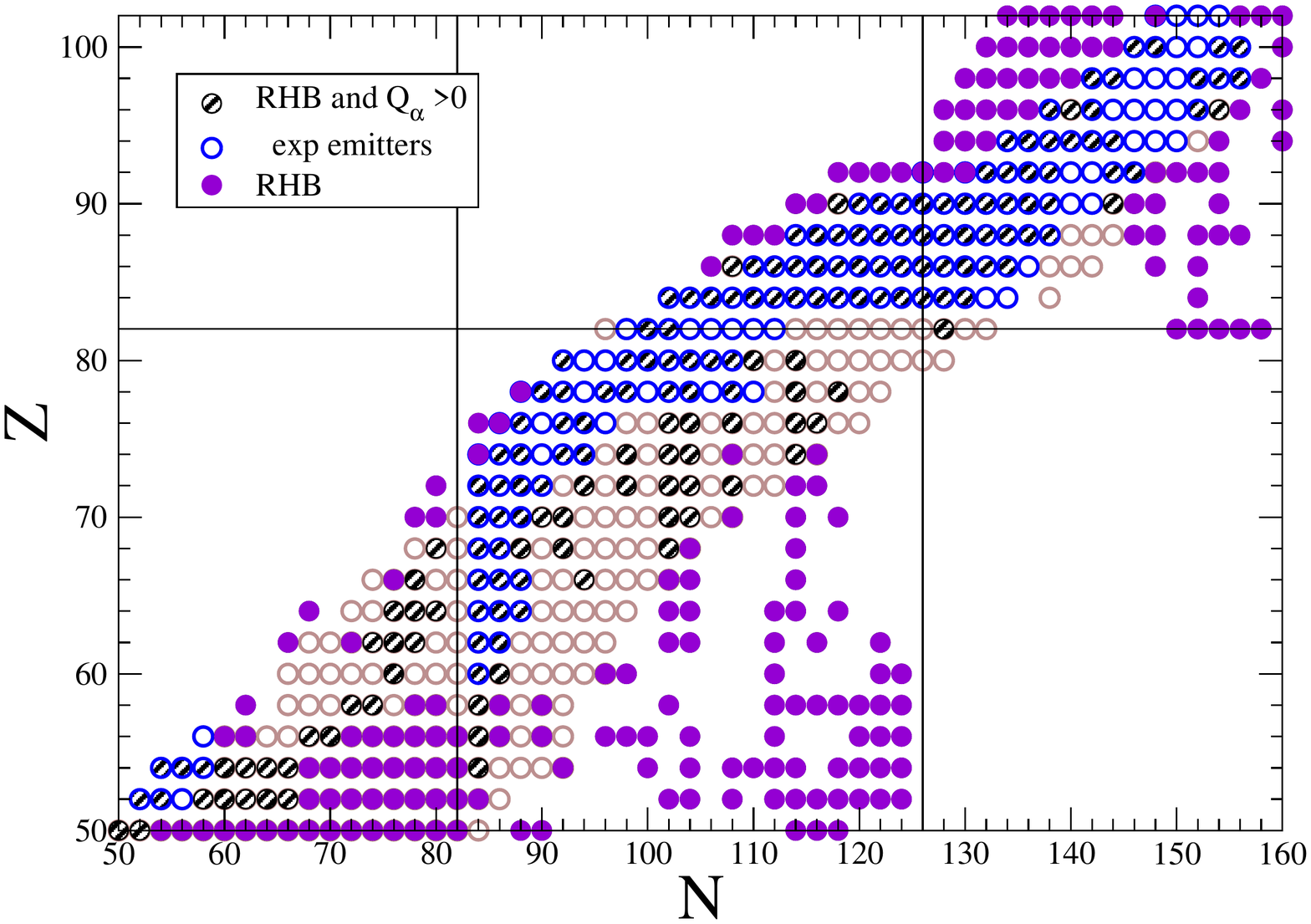}}
 \caption{Experimentally known alpha-emitters for even-even nuclei (in blue), compared to the one predicted to have a small dispersion (see text for details) in their valence state (dashed black) with the RHB calculation, and a positive Q$_\alpha$ from available measured masses. Removing the positive Q$_\alpha$ condition (from available measured masses), allows to predict additional nuclei (full purple dots) with the RHB model}.
 \label{fig:chart}
\end{figure}

The spatial dispersion of valence states is microscopically calculated on the whole nuclear chart of even-even nuclei in the axially-symmetric RHB framework, taking into account pairing and deformation effects.To investigate if spatially localised valence states increase the alpha preformation probability, the neutron or proton dispersion is calculated with the RHB approach (depending on the closest lower N or Z magic number, e.g. proton dispersion for N or Z between 50 and 82, etc.) and the condition (\ref{eq:cond}) is applied, as well as Q$_\alpha >$0, which is a necessary condition for alpha radioactivity. The Q$_\alpha >$0 condition is taken from experimentally measured masses. Fig. \ref{fig:chart} displays such even-even nuclei, compared to the experimentally known alpha-emitters. The present approach  allows to recover a large majority of experimentally known alpha-emitting nuclei: the overall behavior over the nuclear chart is well described, showing that the $n=1$ localisation condition is relevant for alpha-particle emission.  

Several nuclei on Fig. \ref{fig:chart} are however predicted as alpha-emitters, but have not been experimentally tagged so. This could of course be due to a limitation of the general description of alpha radioactivity, into steps initially assuming its localisation and preformation. Another reason could be, that a majority of these nuclei are beta-unstable, and it may be experimentally difficult to look for alpha-emission if the partial beta-decay half-life is several orders of magnitude smaller than the possible alpha one. Hence, these nuclei could be also considered as possible predictions for alpha-emitters, which may have not been detected yet. For instance, it could be interesting to experimentally look for nuclei which are predicted both beta and alpha-emitters around the $^{142}$Gd region. Finally, a last reason could be the role of more elaborate deformations than the axially symmetric one, such as triaxiality and/or octupole deformations, as mentioned in the introduction of section II.

Figure \ref{fig:chart} also displays nuclei having a small spatial localisation from the RHB calculations, as discussed above, removing the Q$_\alpha >$0 condition. This allows to look for possible alpha emitting nuclei, which masses are currently unknown, since the Q$_\alpha >$0 condition was taken from the available measured masses. It shows that, although a few more nuclei for experimentally known masses appear, one of the largest change occurs for exotic nuclei, for which there is no available experimental masses yet. For these very exotic nuclei, detecting alpha disintegration may be challenging, because of the large branching ratio to other decay modes such as beta emission.

Another interesting difference, due to the Q-value condition, deals with intermediate mass nuclei, above N,Z=50: this is due to the frequent occurence of $n=1$ states in these nuclei, compared to heavier one. Hence, the hindrance of alpha emission for these nuclei is largely due to the Q$_\alpha >$0 condition, whereas for heavier nuclei, the delocalisation effect also contributes to the hindrance of alpha emission.
For instance, several nuclei in the Pb isotopic chain, and below, are not predicted to be alpha-emitters, in agreement with the data. This is explained, in the present approach, by the presence of, for instance, the 3s$_{1/2}$ state, having a large spatial extension, due to its $n=3$ value. On the contrary, Sn isotopes can have a large alpha preformation probability, due to the presence of $n=1$ states in these lighter nuclei. However, most of them have a negative  Q$_\alpha$ value, and hence, cannot be detected as alpha-emitters. These effects shall be discussed in more details in the next section, with the comparison of the $^{104}$Te and $^{212}$Po cases.

\section{Role of localisation}

In order to investigate more precisely the role of spatial localisation on alpha preformation and emission probability, two benchmark cases are compared: $^{212}$Po and $^{104}$Te. The former is a well-known alpha-emitter, whereas the latter has recently been evidenced as an alpha emitting nucleus \cite{aur18,xia19}. Indeed, the deduced alpha preformation probability has been found larger than in $^{212}$Po. Moreover, the alpha lifetime of $^{104}$Te has also been recently described by an alpha+core approach \cite{ibr19}, showing the relevance of connecting alpha cluster approaches to alpha emission.

\begin{figure}[tb]
\scalebox{0.29}{\includegraphics{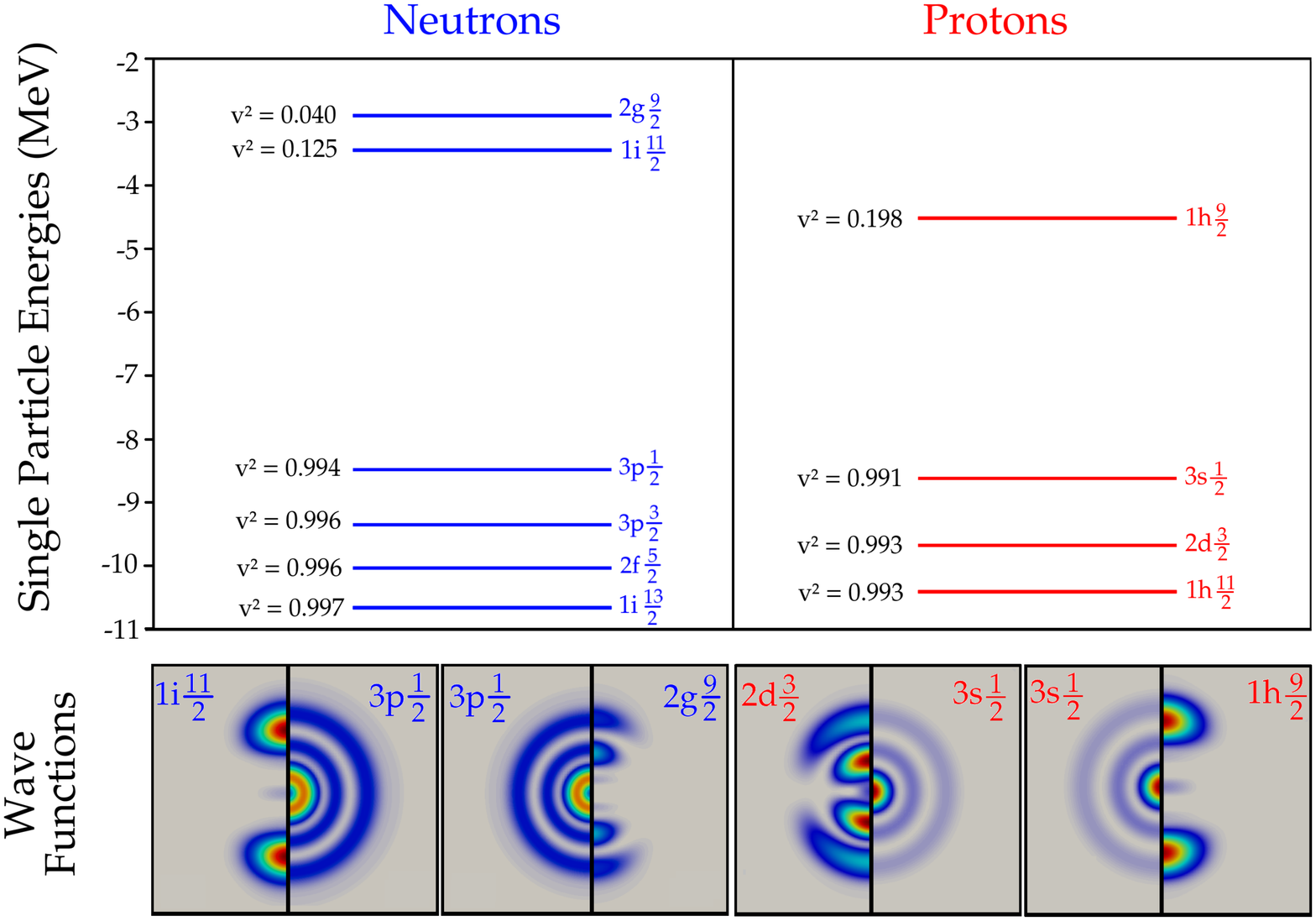}}
 \caption{Spectrum and partial densities predicted for $^{212}$Po with the RHB calculations. The states are Kramers (Nilsson) states, with a degeneracy of 2, but are displayed with the quantum numbers of the spherical state from which they are stemming.}    
 \label{fig:212Po}
\end{figure}

Figure \ref{fig:212Po} displays the single-particle spectrum of $^{212}$Po obtained with the RHB calculations. Because of the pairing effect, the occupation probability of each state is indicated. For the valence neutrons states, not only $n=1$ states are involved, but also the 2g$_{9/2}$ state, which increases the spatial dispersion, because it has $n=2$ \cite{ebr18}. The contribution to the spatial localisation is detailed on the bottom of the figure, displaying the partial densities obtained from RHB calculations. The 1i$_{11/2}$ state, although of large $\ell$ value, is much more localised than both the 2g$_{9/2}$  and 3p$_{1/2}$ states, in agreement with our main point on spatial localisation (namely $n$ dependence but $\ell$ independence). It should be noted that the vicinity of the 2g$_{9/2}$ state with the 1i$_{11/2}$ one, is due to the pseudo-spin symmetry (PSS) \cite{gin99}, which plays an important role to describe the behavior of nuclei in this region of the nuclear chart \cite{ebr16}.

In the case of the protons, the valence state of $^{212}$Po is almost only made of the 1h$_{9/2}$ state. It is also more localised than states with larger $n$, as seen on the partial densities. Table I summarises the respective dispersions calculated for the valence states of $^{212}$Po, showing the decisive role of the n quantum number: a $n>$1 value drastically increases the dispersion, by about a factor 2 or more: the localisation effect is decreased due to the presence of $n=2,3$ states in the valence region. This has two roots: i) being a heavy nucleus, $n>$1 states are likely to contribute to the single particle spectrum ii) PSS imposes the vicinity of a $n=2$ state close to the 1i$_{11/2}$ neutron valence state. It is therefore expected that in lighter nuclei such as $^{104}$Te, the blurring of the spatial localisation disappears because of the absence of $n>$1 states. The dispersion is also expected to be smaller because of its A dependence (see Eq. \ref{eq:res}).

\begin{table}[h!]
\begin{center}
\setlength{\tabcolsep}{4pt}
\begin{tabular}{ |c| c | c | c || c|  c |c|}
\hline
 $^{212}$Po&  2g$_{9/2}$ &1i$_{11/2}$ &3p$_{1/2}$ & 1h$_{9/2}$&3s$_{1/2}$ &2d$_{3/2}$ \rule[-7pt]{0pt}{20pt}\\
\hline
$\Delta$r (fm) & 1.95 &1.17 &2.40 &1.08 &2.45 &1.91  \rule[-7pt]{0pt}{20pt} \\
\hline
\end{tabular}
\caption{Spatial dispersion of neutron (left part) and proton (right part) single-particle states of $^{212}$Po, calculated with the RHB approach.}
\end{center}
\end{table}

$^{104}$Te is therefore a specifically interesting nucleus, to study alpha preformation probability. It belongs to the lightest region where Q$_\alpha$ remains positive. Moreover, this N=Z nucleus would also correspond, in a simple picture, to an alpha particle on top of a doubly magic core. However, this nucleus is close to the proton drip-line, making its description delicate. In the present approach, the axially deformed RHB calculation finds its ground state with a small deformation ($\beta_2$=0.14) with a proton valence state at a slightly positive energy, by 140 keV. This could be due to a limitation of the model to describe nuclei close to the drip-line, as mentioned above. However, due to the Coulomb barrier, the static description of this proton quasi-bound state, could still be considered as valid: our main goal is to focus on the spatial localisation of the wave functions, and not to study the particle emission process itself. We also wish to consider a global approach, such as the relativistic Hartree-Bogoliubov, rather than using more dedicated models to accurately describe a given set of nuclei. Finally, it should be noted that the output of the RHB calculations, also shows a collapse of the pairing effect. This result favors a description of $^{104}$Te  as a $^{100}$Sn core + 4 valence nucleons, compared to a case with pairing, where the $^{100}$Sn states and the valence ones would be mixed, with occupation factors between 0 and 1.

\begin{figure}[tb]
\scalebox{0.29}{\includegraphics{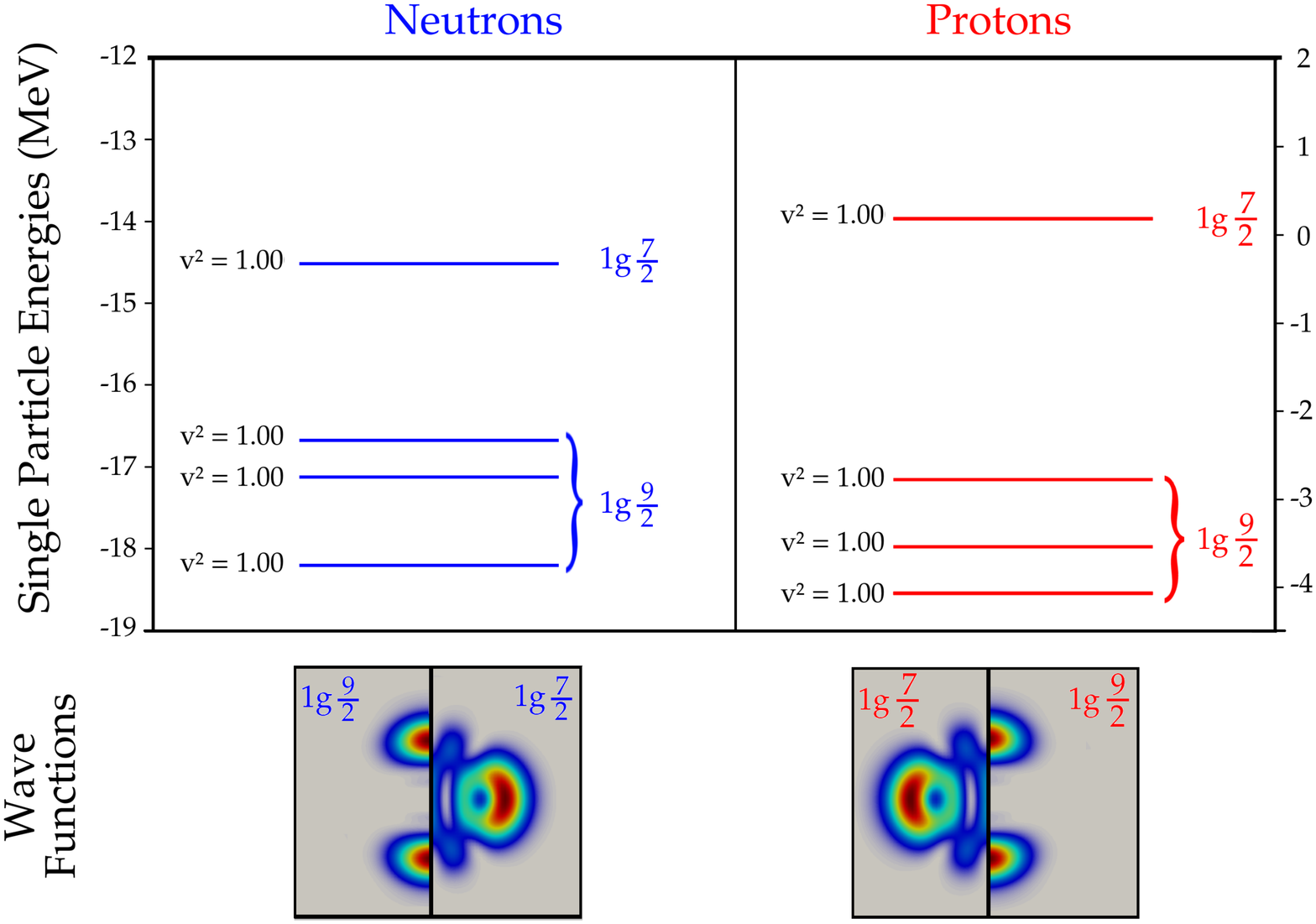}}
 \caption{Same as Fig. \ref{fig:212Po} for $^{104}$Te}    
 \label{fig:104Te}
\end{figure}

Fig \ref{fig:104Te} shows that only $n=1$ states are involved as valence states, namely the 1g$_{7/2}$ state both for neutrons and protons: compared to  $^{212}$Po, $^{104}$Te is closer from the lightest nuclei, where clusters states can be found. The corresponding partial densities, as well as the one of the 1g$_{9/2}$ state, are spatially localised, although the 1g$_{7/2}$ state shows more extension than the 1g$_{9/2}$. In addition $^{104}$Te being a lighter nucleus than $^{212}$Po, the dispersion of the $n=1$ state is also smaller: for instance, 
the 1g$_{9/2}$ one is 0.98 fm, to be compared with the values for $n=1$ in Table I. It should be noted that the dispersion of the 1g$_{7/2}$ state is about 1.5 fm, which is larger, probably due to the difficulty to describe such a nucleus close to the drip-line, involving quasi-bound states: the dispersion of the 1g$_{9/2}$ state is more representative of the typical spatial dispersion at work in this nuclei. Under this assumption, the neutron valence states of $^{212}$Po have a spatial dispersion in average, about 40 \% larger than the 1g$_{9/2}$ of $^{104}$Te. 

It should be noted that the lowest neutron Kramers (Nilsson) states, originating from the spherical 2d$_{5/2}$ state in $^{104}$Te, are located at -11.2 MeV, showing that the degeneracy raising between PSS partner states \cite{alb02} is much larger than in $^{212}$Po. This is due to the 
effect of deformation in $^{104}$Te, overcoming the one of the PSS.

\section{Study of the preformation factor}

It could be useful to study the impact of spatial localisation of the valence state, on the alpha preformation factor. First, the localisation criterion (\ref{eq:cond}), calculated microscopically, shall be tested using the data on the alpha emission half-life (through a phenomenological fit-like formula). This shall also allow to study the impact of shell effects on the alpha preformation factor. Finally, relating the alpha preformation probability to the radial quantum number of the valence state, can be of interest. On this purpose, a phenomenological law shall be derived.

\subsection{Evaluation of the alpha preformation factor}

\subsubsection{Evaluation of the localisation criterion}

In order to substantiate the above findings, it could be useful to consider a complementary approach, and evaluate the alpha preformation factor. 
A relevant way is to consider the following formula for the alpha emission half-life:

\begin{equation}
Log_{10}T^{Pheno}_{1/2}(s) =\frac{9.54(Z-2)^{0.6}}{\sqrt{Q_\alpha}}-51.37
\label{eq:br}
\end{equation}

where Q$_\alpha$ is in MeV. In \cite{bro92}, it has been shown that Eq. (\ref{eq:br}) both accurately describes the experimental data and compares well to theoretical WKB approximation inferring a preformation probability P=1. Therefore, discrepancies of the data with respect to this formula shall be driven, as a first approximation, by the behavior of the alpha preformation factor, although other effects could play a role, such as triaxal or octupole deformation, resonance densities or description of the alpha formation with respect to the barrier penetration \cite{tan20}.
More precisely, the ratio  T$^{Pheno}$/T$^{Exp}$=W$^{Exp}$/W$^{Pheno}$, where W is the total $\alpha$ emission probability, shall give an evaluation of the alpha preformation factor, as discussed in \cite{bro92}.

\begin{figure}[htb!]
\scalebox{0.33}{\includegraphics{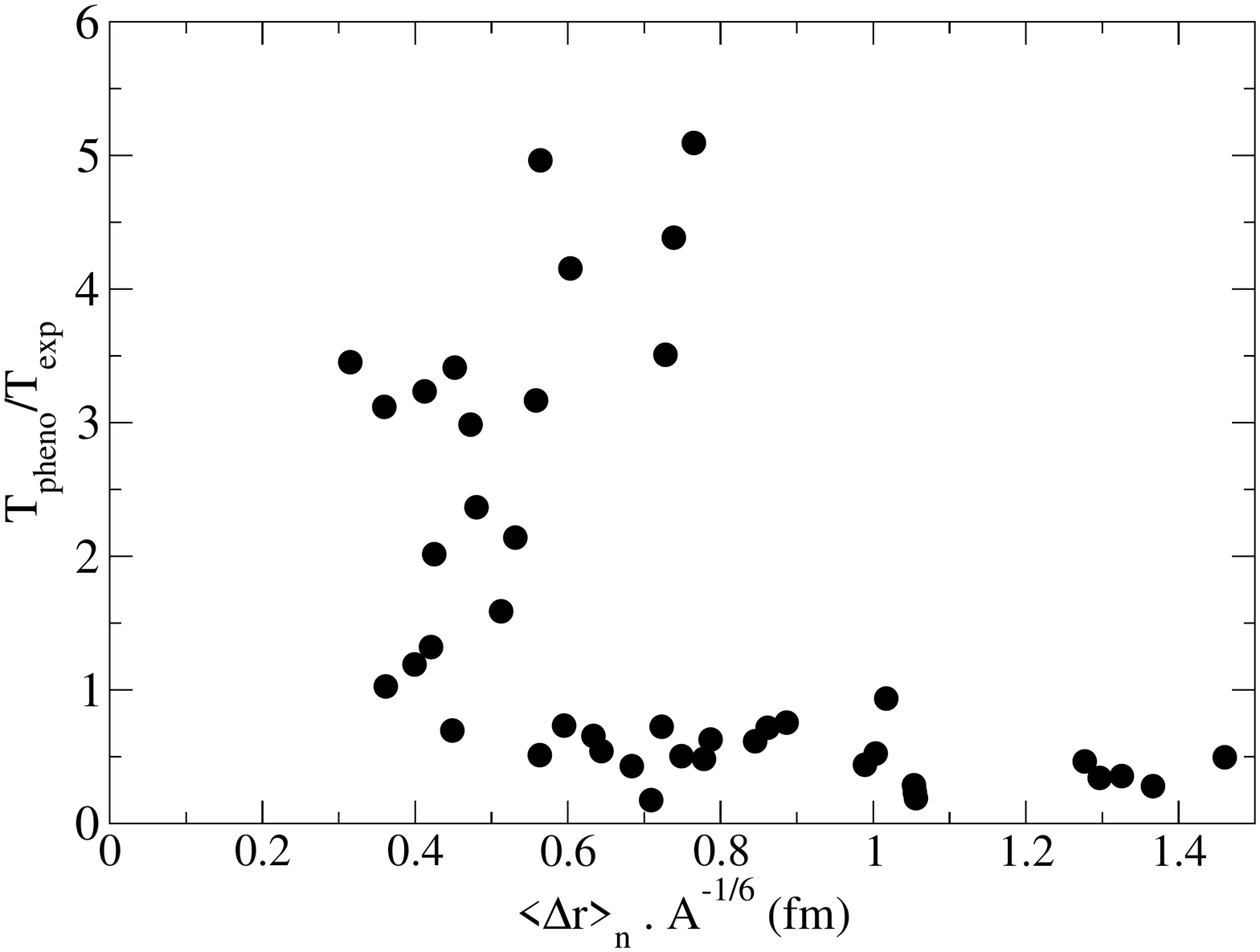}}
 \caption{Ratio of the phenomenological to experimental alpha emission half-life for Z=84,86,88 even-even nuclei as a function of the average dispersion of the neutron valence state (see Eq. \ref{eq:cond} for details) }    
 \label{fig:ratiodel}
\end{figure}

In order to provide a quantitative relation between the impact of the localisation criterion of Eq. (\ref{eq:cond}) and the alpha formation probability, Fig. \ref{fig:ratiodel} displays the T$^{Pheno}$/T$^{Exp}$ ratio as a function of the average dispersion of the neutron valence state, for the Z= 84, 86 and 88 isotopes. Large alpha preformation factors (T$^{Pheno}$/T$^{Exp}\gtrsim$1) involve small average nucleonic dispersion of the neutron valence states ($<\Delta r>.A^{-1/6}\lesssim$ 0.7 fm), whereas small alpha preformation factors (T$^{Pheno}$/T$^{Exp}\lesssim$1) can involve large average nucleonic dispersion of the neutron valence states ($<\Delta r>.A^{-1/6}$ up to 1.4 fm). This shows the relevance of the localisation criterion  (\ref{eq:cond}) on the alpha decay phenomenon.

 We have also performed a systematic calculation, showing that almost all the nuclei with the smallest preformation factor are odd ones, in agreement with the hindrance effect of the alpha preformation, known to occur in such nuclei \cite{vio66,san15}. In the case of even-even nuclei, out of 140 nuclei where the T$^{Pheno}$/T$^{Exp}$ ratio has been calculated, 56 are predicted as localised by the RHB calculation. The mean value of the T$^{Pheno}$/T$^{Exp}$ ratio is 2.4$\pm$0.2 for these nuclei, to be compared to 1.3$\pm$0.2 for the nuclei which do not fulfill the localisation criterion. Although there is a rather small difference between these two values, it is significant enough to consider that the present quantity is a relevant probe, at least on the qualitative level, for the alpha preformation factor in nuclei. However, it should be noted that reducing alpha emission phenomenon to the occurence of localised valence state, is of course an approximation, and a one-to-one quantitative correlation is not expected.

\subsubsection{Analysis of the shell effects}

\begin{figure}[htb!]
\scalebox{0.33}{\includegraphics{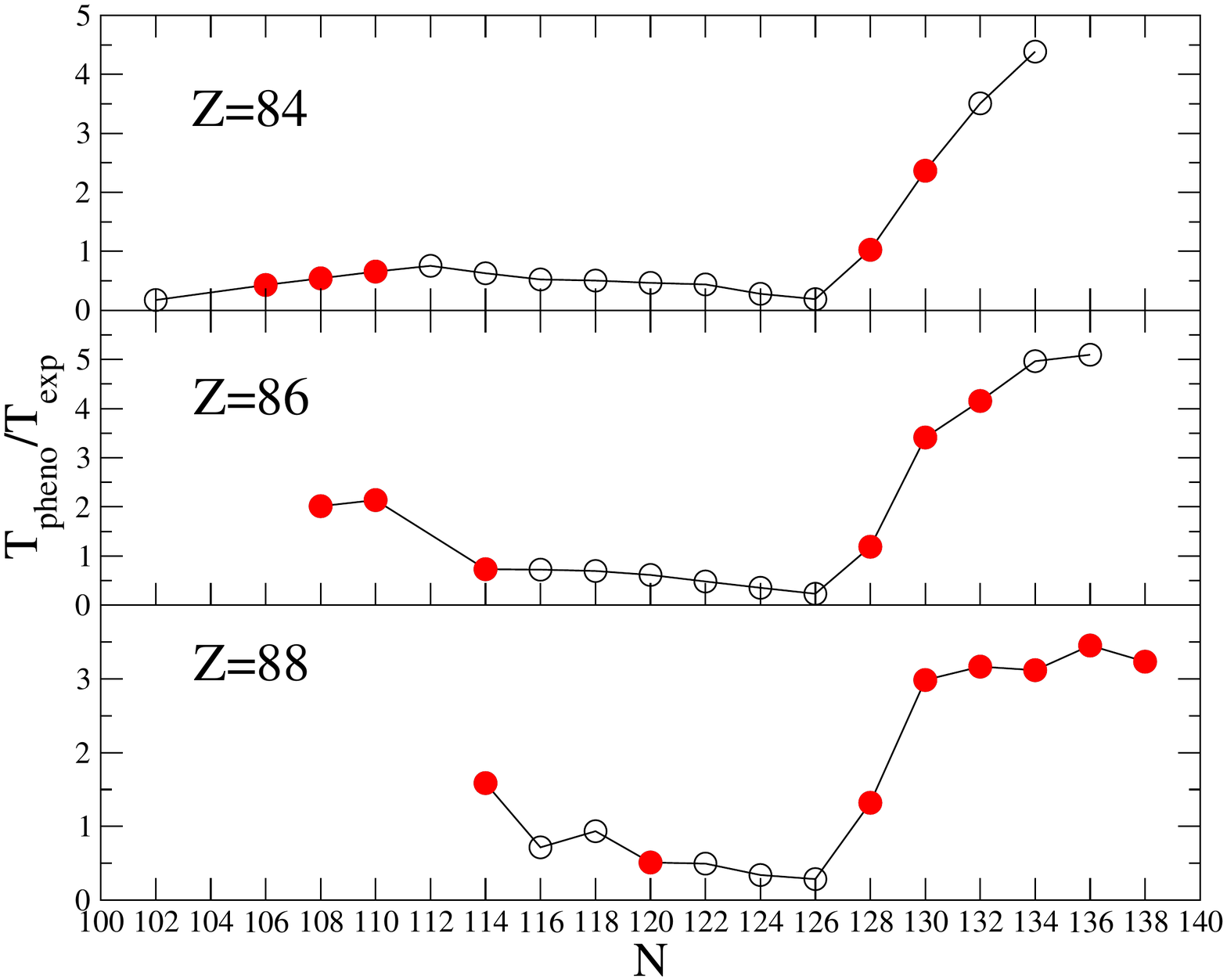}}
 \caption{Ratio of the phenomenological to experimental alpha emission half-life for Z=84,86,88 even-even nuclei. Those predicted with a small dispersion (see text) are displayed in red.}    
 \label{fig:ratio}
\end{figure}

The above mentioned ratio can also be used to study the impact of shell effects on alpha decay.
Fig. \ref{fig:ratio} displays the T$^{Pheno}$/T$^{Exp}$ ratio for Z$\gtrsim$82 nuclei, where the experimental data is known on relatively large isotopic chains. It should be reminded that this ratio shall scales the alpha preformation factor (see e.g. Fig. 4 of  \cite{bro92}). Only even-even nuclei are displayed. Fig. \ref{fig:ratio} first shows that there is a sharp increase of this ratio for N$\ge$128, exhibiting an important shell effect. Nuclei which are predicted as localised ones (Eq. (\ref{eq:cond})), both for the protons and neutrons valences wave functions, from the microscopic RHB calculation, are displayed in red. They correspond to most of nuclei having the largest alpha preformation factor, especially around the N=128 shell closure. This shows a correlation between localisation and alpha preformation probability, and gives quantitative grounds to the present analysis of the role of localisation on the alpha preformation factor.

The present description also allows for a more detailed discussion. The N=126 shell effect is phenomenologically known to impact the alpha preformation probability \cite{bro92}. 
 In the present analysis, the strong increase of the preformation probability, starting from N=128, comes from the filling of the 1i$_{11/2}$ state (although partially, i.e. together with the 2g$_{9/2}$ state, as seen on Fig. \ref{fig:212Po}) whereas the regular decrease of the preformation probability before N=128 is related to the progressive filling from $n=1$ state (1i$_{13/2}$) to n=2,3 states such as 2$_{5/2}$, 3p$_{3/2}$ and 3p$_{1/2}$.
  
The case of the smaller Z=82 shell effect, on the preformation probability \cite{and13}, is more complicated to study: looking to the experimental alpha-emitters (Fig. \ref{fig:chart}), an isotopic chain of alpha-emitters nuclei through N=126 involves many nuclei which are moreover close to the stability line (i.e. just above $^{208}$Pb). Hence, the filling of the $n=1$ state beyond N=126 happens in a similar way for the various isotopic chains, from Z=84 (Po) to Z=92 (U). On the contrary, following an isotonic chain of alpha-emitters nuclei through Z=82, involves much fewer nuclei, due in part to the proton drip-line: more intense nuclear structure effects are expected, and the occurrence of the $n=1$ state just above Z=82 is expected to be less systematic. 

\subsection{Phenomenological alpha preformation model}

Due to the complexity of the properties of the alpha emission \cite{del10}, phenomenological models are often used, as illustrated by the successful Geiger-Nuttall law \cite{gei11}. Following this spirit, it could be finally useful to provide, as an alternative way, a phenomenological relation between the alpha preformation probability and the spatial localisation, over the nuclear chart. This law is expected to give only order of magnitudes, due to the simplicity of the approach, but relating the alpha preformation probability to the radial quantum number of the valence state can be of interest.

The microscopic RHB calculations of the spatial dispersion of the valence states of $^{212}$Po (see Table I) and $^{104}$Te, allows to calculate their average value, leading to the following relation, as discussed in the previous section:

\begin{equation}
<\Delta r(^{212}Po)>\simeq 1.4<\Delta r(^{104}Te)>
\label{inp}
\end{equation}

Inspired by the form of the Geiger-Nuttall law, our ansatz, for the alpha preformation probability P, is a power law as a function of the localisation parameter:
 
\begin{equation}
P=10^{-B\alpha_{loc}+C}
\label{ans}
\end{equation}

where B and C are constants to be determined. In a recent experiment \cite{aur18}, the alpha preformation probability in $^{104}$Te was deduced to be at least 3 times larger than in $^{212}$Po.
We therefore take, as typical orders of magnitudes, P($^{104}$Te)=1 and P($^{212}$Po)=0.1 to mimic this effect. This allows to determine B and C, namely with Eq. (\ref{eq:res}):

\begin{equation}
Log \sqrt{P} = 1-\left(\frac{A}{100}\right)^{1/6}\sqrt{2n-1}
\end{equation}  

Since the above dependence on A is rather weak over the nuclear chart, this law can be approximated by 

\begin{equation}
Log \sqrt{P}\simeq 1-\sqrt{2n-1}
\label{pre}
\end{equation}

where $n$ is the radial quantum number of the valence state of the considered nucleus. In the case of pairing effect, $n$ can be taken as the average of the $n$ values of the valence states, weighted by their occupation probabilities. The alpha preformation probability, calculated with Eq. (\ref{pre}), is reduced by a factor 20-30 when the valence state switches from a $n=1$ to a $n=2$ state. This result is in agreement with the typical observed variation of the estimation of the preformation of the alpha probability extracted from the data (see e.g. Fig. 4 of Ref. \cite{bro92}). It should be noted that
taking the upper limit  P($^{212}$Po)=0.3, instead of 0.1, would lead to a reduction of a factor 10 of the alpha preformation probability, from a $n=1$ to a $n=2$ state. Considering that approximations are done on an exponential quantity, the present approach should remain qualitative.

\section{Conclusion}

The alpha preformation probability has been analysed through the behavior of the spatial localisation of nucleonic states. A criterion for localised states has been first established.
The systematically more localised $n=1$ states, independently of the orbital angular momentum value, allow to pinpoint nuclei which are more likely to have a large preformation probability over the nuclear chart.  In order to compare with experimentally known alpha-emitters, axially deformed RHB calculations have been performed over the nuclear chart to provide microscopic spatial dispersions. The systematic occurrence of more localised valence states (which do have $n=1$) shows a pattern which is in agreement with experimentally known alpha-emitters. The investigation of the single-particle spectra of $^{212}$Po and $^{104}$Te, allows to understand in more details why the alpha preformation probability is larger in the latter than in the former. This is partly due to the PSS symmetry at work in $^{212}$Po, involving a n=2 state, and to the fact that, being a lighter nucleus, $^{104}$Te involves almost only $n=1$ states, each of them also having a bit smaller dispersion due to the mass effect on the localisation parameter. 

In order to  study in a complementary way these results, a phenomenological evaluation of the alpha preformation factor has been undertaken. It confirms the relevance of the criterion for localised states, and also shows that it is correlated to the enhancement of the alpha preformation factor, especially after shell closure. Finally, a phenomenological law relating the preformation probability to the radial quantum number of the valence states, has been extracted.
  
 All these results show the relevance of relativistic approaches, not only to describe cluster states in nuclei, but also to grasp the main properties of alpha radioactivity, as also shown in Ref. \cite{mer20}. The present approach does not aim to be very accurate, especially in the difficult domain of alpha radioactivity, where various orders of magnitudes are at stake. Effects of more advanced deformations could be studied, such as the role of triaxiality and/or octupole deformations. The description of $^{104}$Te could also be improved with a more dedicated model, suited for nuclei close to the drip line. Finally, the building of a four-body alpha wave function, together with quartetting correlations, could be considered in a near future.

\begin{acknowledgments}
This publication is based on work supported in part by the framework of the Espace de Structure et de r\'eactions Nucl\'eaires Th\'eorique (ESNT esnt.cea.fr) at CEA.
\end{acknowledgments}

\end{document}